\documentclass[conference]{IEEEtran}
\usepackage{booktabs}
\usepackage{multirow}
\usepackage{tabularx}
\usepackage{array}
\usepackage{graphicx}
\usepackage{svg}
\usepackage{xcolor}
\usepackage{subcaption}
\usepackage{amsmath}
\usepackage{amssymb}
\usepackage[normalem]{ulem}
\usepackage{comment}

\AtBeginDocument{%
  \setlength{\textfloatsep}{5pt plus 2pt minus 2pt}
  \setlength{\dbltextfloatsep}{6pt plus 2pt minus 2pt}
  \setlength{\floatsep}{5pt plus 2pt minus 2pt}
  \setlength{\intextsep}{5pt plus 2pt minus 2pt}
  \setlength{\abovecaptionskip}{2pt}
  \setlength{\belowcaptionskip}{0pt}%
}
\ifCLASSINFOpdf
\else
\fi

\begin{document}
%
\title{\textit{\textbf{NIFA}}: \uline{\textbf{N}}onlinear \uline{\textbf{I}}MC enhanced \uline{\textbf{F}}PG\uline{\textbf{A}} for efficient ML inference
\vspace{-0.2cm}
}

\author{%
\IEEEauthorblockN{Jiajun Hu\textsuperscript{1}, Ruthwik Reddy Sunketa\textsuperscript{1}, Lei Zhao\textsuperscript{2}, Archit Gajjar\textsuperscript{2}, Luca Buonanno\textsuperscript{2}, Aman Arora\textsuperscript{1}}
\IEEEauthorblockA{\textsuperscript{1}Arizona State University, Tempe, AZ, USA \quad \textsuperscript{2}Hewlett Packard Enterprise Labs, Fort Collins, CO, USA}
\IEEEauthorblockA{\{jiajunh5, rsunketa, aman.kbm\}@asu.edu \quad \{lei.zhao, archit.gajjar, luca.buonanno\}@hpe.com}
}

\maketitle

\begin{abstract}
Recent FPGAs have improved deep learning (DL) inference efficiency by introducing tensor blocks and enabling in-BRAM computation. ReRAM-based analog in-memory computing (IMC) cores offer an order of magnitude higher compute density and energy efficiency than conventional digital computation by performing vector-matrix multiplication (VMM) directly within the ReRAM crossbar. Prior work has integrated such IMC blocks into FPGAs for DL inference. However, conventional IMC designs support only static-weight VMM, while nonlinear and dynamic matrix-matrix multiplications (DIMM) are still handled by the FPGA fabric. As a result, the benefits of IMC are largely limited to static-weight DL models, whereas Transformer-based models, which require frequent nonlinear and DIMM operations, achieve only limited benefit. In addition, ADCs within the IMC block consume more than 70\% of area and power, further limiting system efficiency and scalability. To address these issues, we propose a novel FPGA architecture that integrates an ADC-free IMC alternative into FPGAs, replacing the conventional ADC with analog content-addressable memories (ACAMs) that natively perform nonlinear operations inside the IMC block. To fully utilize this new block, we conduct an FPGA-aware design-space exploration that determines the optimal crossbar sizes while balancing FPGA area, flexibility, and DL performance. We further deploy an efficient mapping that uses ACAMs to efficiently perform DIMM operations, extending architectural applicability to Attention computation. Across CNN and Transformer-based benchmarks, our proposed FPGA architecture achieves up to $40\times$ and $1.9\times$ higher energy efficiency, and $4.1\times$ and $2.5\times$ area efficiency. Overall, the proposed architecture significantly improves the FPGA DL inference efficiency and shows robust efficiency gain on Transformer-based workloads across long input sequences, advancing domain-specialized FPGA design.

\end{abstract}


%
\IEEEpeerreviewmaketitle

\vspace{-2mm} 
\section{Introduction}


As deep learning (DL) has been widely adopted across modern applications, recent FPGA architectures have begun integrating domain-specific blocks to improve DL inference efficiency. A natural target is \textit{vector-matrix multiplication} (VMM), the dominant operation in DL workloads: early works demonstrated that embedding a dedicated \textit{matrix multiplier} block directly into the FPGA fabric yields substantial efficiency gains \cite{tensorslice,sparsetensorslice}. FPGA vendors have since followed suit, incorporating \textit{AI tensor blocks} into commercial devices, validating the effectiveness of this approach~\cite{stratix10}.
Researchers have also proposed in-memory compute (IMC) architectures for FPGAs to reduce on-chip data movement through the routing fabric and increase FPGA compute density \cite{comefa,bramac,m4bram,ccb}.
While these works show 
speedups upto 3x for multiple DNN workloads, the energy consumption remains high.

\begin{figure}[!t]
\centering
\includegraphics[width=0.45\textwidth]{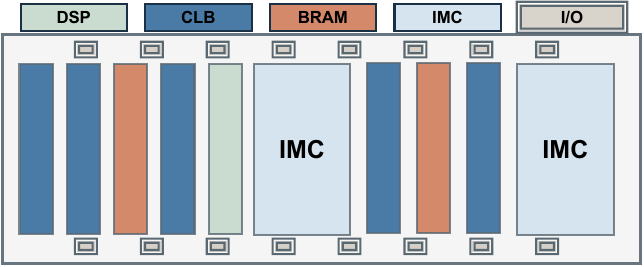}
\caption{\small Overview of the proposed IMC-enhanced heterogeneous FPGA architecture. IMC hard blocks are embedded as dedicated columns alongside CLBs, DSPs, and BRAMs.}
\label{fig:dpe_fpga_overview}
\vspace{-0.3cm}
\end{figure}

Unlike prior in-BRAM compute approaches, recent work has proposed integrating resistive RAM (ReRAM)-based \textit{Analog In-Memory-Compute} core directly into the FPGA fabric, achieving an order-of-magnitude efficiency improvement on  inference of convolutional neural network (CNN) workloads ~\cite{azurelily,Analog_in_mem}. 
Such analog IMC blocks exploit Kirchhoff's laws to perform VMM operations directly inside the memory, delivering significantly higher compute density and energy efficiency than digital computation. Fig.\ref{fig:dpe_fpga_overview} shows an overview of the IMC-enhanced FPGA architecture, where IMC blocks are embedded as dedicated fabric columns and connected to the global routing network. All inter-tile communication occurs in the digital domain; digital-to-analog and analog-to-digital conversion is handled internally within each IMC tile.

These efficiency gains are well demonstrated in CNN inference, where layers primarily involve multiplying statically programmed weights with feature maps \cite{analogue_compute,PRIME}. In contrast, modern Transformer-based \cite{transformer} workloads exhibit fundamentally different characteristics.
Specifically, the Attention mechanism requires dynamic input matrix multiplication (DIMM), in which both operands are computed at runtime. Additionally, Attention involves multiple nonlinear operations, such as exponentiation and activation, that are not natively supported by prior IMC blocks.
The computational complexity of Attention scales quadratically with sequence length, further amplifying its cost. In existing IMC approaches, the entire Attention pipeline typically falls back to FPGA soft logic, negating the benefits of integrated IMC acceleration. As sequence length increases, Attention increasingly dominates overall computation, emerging as a critical bottleneck that cannot be alleviated simply by adding more IMC blocks.


To address this gap, we propose \textit{NIFA}, an FPGA architecture integrates \textit{NL-DPE} block into FPGA fabric \cite{nldpe}. The \textit{NL-DPE} block 
integrates a ReRAM based crossbar with analog content-addressable memories (ACAM) and no ADCs.
The block provides native support for nonlinear functions such as exponentiation and activation, and delivers up to 30$\times$ energy efficiency than conventional ADC-based IMC blocks \cite{nldpe,raceit}. 
Combined, these features significantly improve the performance and energy-efficiency for Transformer-based workloads, reducing soft-logic based computation significantly. 
This paper makes the following contributions:
\begin{itemize}
    \item We propose \textit{NIFA}, a novel IMC-enhanced FPGA architecture that provides in-block nonlinear functionality support and shows up to 40$\times$ higher energy efficiency and 1.7$\times$ higher throughput efficiency than SOTA for end-to-end DL benchmarks.
    \item We conduct an FPGA-aware design space exploration (DSE), which quantitatively evaluates the tradeoff between IMC block size, FPGA composition (percentage of FPGA area occupied by IMC blocks), DL performance, and FPGA flexibility using representative DL micro-benchmarks and non-DL benchmarks.
    \item We present the first work that demonstrates using IMC blocks on FPGA for DIMM in Transformer. We deploy an efficient mapping that accelerates Attention computation in the log domain, combining the FPGA flexibility and IMC's efficient in-block nonlinear functionality, demonstrating 1.7$\times$ performance efficiency than SOTA on BERT-Tiny on long sequence length.
\end{itemize}







\vspace{-2mm} 
\section{Related Work}


FPGA vendors have long integrated DSP slices and Block RAMs to improve FPGA performance for common workloads. Recent studies propose incorporating domain-specific hard blocks into FPGAs to further improve DL throughput. \textit{Hamamu}~\cite{hamamu} and \textit{Tensor Slices}~\cite{tensorslice} replace a portion of programmable logic with hardened matrix multipliers that support multiple modes and precisions.
\textit{Systolic Sparse Tensor Slices}~\cite{sparsetensorslice} further extend this idea to accelerate structured sparse workloads. While these approaches improve compute throughput, the compute density of an FPGA still remains relatively low compared to ASICs, and the data movement to shuttle operands between RAM and compute blocks through the global routing leads to significant energy consumption. 


To mitigate these bottlenecks, some work embeds compute units into BRAMs. \textit{CoMeFa}~\cite{comefa} augments BRAMs with bit-serial processing elements at the sense amplifier outputs, enabling in-BRAM computation without external data movement. \textit{BRAMAC} and \textit{M4BRAM}~\cite{bramac, m4bram} add compact dummy arrays and customized ALUs within the BRAM tile to support MAC operations at mixed precision. All three achieve modest throughput improvements on DL benchmarks by exploiting the parallelism inherent in wide BRAM arrays. 

Compared to digital IMC, ReRAM-based analog IMC offers an order-of-magnitude higher compute density and energy efficiency by exploiting physical mechanisms such as Kirchhoff's law to perform computation directly within the memory. \textit{Azure-Lily}~\cite{azurelily} integrates such an analog IMC block into an FPGA fabric, demonstrating $6.58\times$ latency reduction and $8{,}741\times$ energy efficiency improvement over CLB/DSP-only implementations on CNN benchmarks. Modern DL benchmarks such as Transformer based networks and LLMs are not evaluated. The in-block ADC arrays, which consume over 70\% of the block area and energy, limit the system-level scalability and efficiency. Furthermore, the design-space exploration  is limited to the block level and FPGA integration evaluation such as area budget and flexibility-generality tradeoffs are not evaluated. 

Across all three lines of prior work, no existing FPGA hard block provides native in-block support for nonlinear functions, which must fall back to CLBs, creating a throughput bottleneck, especially for modern Transformer-style workloads. 
This work, NIFA, integrates an IMC block which supports native in-block nonlinear computation into FPGA, while evaluating the benefits this feature provides for modern Transformer-based workloads. 
We also perform a two-round FPGA-aware DSE that systematically quantifies the tradeoff among DL throughput, area budget, and general-purpose flexibility.

\vspace{-2mm} 
\section{Background}

\begin{figure}[t]
\centering
\includegraphics[width=0.95\columnwidth]{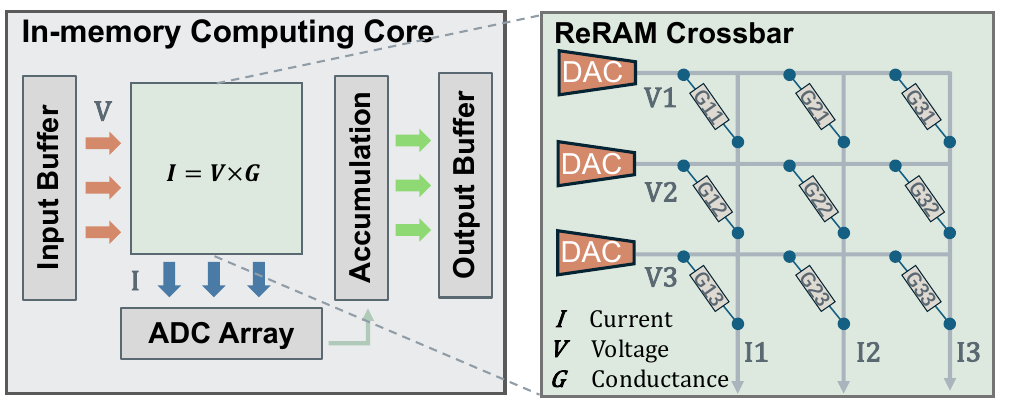}
\caption{\small Example ReRAM-based IMC dot product engine performing VMM: $I = V\times G$. V1, V2 and V3 are input voltage vector applied to each row. I1, I2 and I3 are resultant current accumulated in each column.} 
\label{fig:imc_core}
\vspace{-0.3cm}
\end{figure}

Fig.~\ref{fig:imc_core} shows a ReRAM-based IMC core, which consists of input/output buffers, a ReRAM crossbar array, accumulation logic, and ADC. The crossbar stores DNN weights as ReRAM conductances and performs VMM in the analog domain. The input vector is  applied to the rows, and the accumulated current at each column output is the dot product result. In this figure, the output current in the first column is computed: $I1=V1\times G11+V2\times G12+V3\times G13$. 
To avoid DACs at the inputs, a bit-slicing technique is used: the inputs are decomposed into single-bit slices, each fed into the crossbar serially, and the partial results are accumulated to reconstruct the full-precision output. In this example, the first bits of $V1$, $V2$, $V3$ are fed through the DAC and crossbar, then accumulated with the following bits in the accumulation buffer and output. For N-bit precision, this requires N slices to be fed sequentially. However, within the IMC core, each stage is independent and operates in a pipelined fashion. 
Analog computation is subject to device noise~\cite{mao_crossbar_noise}. Noise-aware training and calibration recover baseline accuracy under realistic variation; NIFA adopts \textit{Azure-Lily}'s noise model~\cite{azurelily}.

\vspace{-2mm} 
\section{Proposed System}
Recent IMC research has proposed an ADC-free IMC block, \textit{NL-DPE}, with native nonlinear computation capability~\cite{nldpe,raceit,zhao_noise_finetune} and shows great efficiency compared to normal ADC-based IMC blocks. 
We integrate this block into the FPGA fabric as a first-class hard block, unlocking its nonlinear functionality for CNN and Transformer workloads.
ReRAM and the FPGA's CMOS logic are fabricated in different layers and do not interfere: the ReRAM cells are a back-end-of-line (BEOL) deposit in the metal interconnect above the transistors, decoupled from the front-end transistor node. The FPGA's programmable fabric can therefore remain at the leading-edge CMOS node while the ReRAM cells occupy the metal layers above, incurring no logic-density penalty. 

\subsection{NL-DPE Block Architecture}
\label{sec:nldpe_block}
\begin{figure}[t]
\centering
\includegraphics[width=\columnwidth]{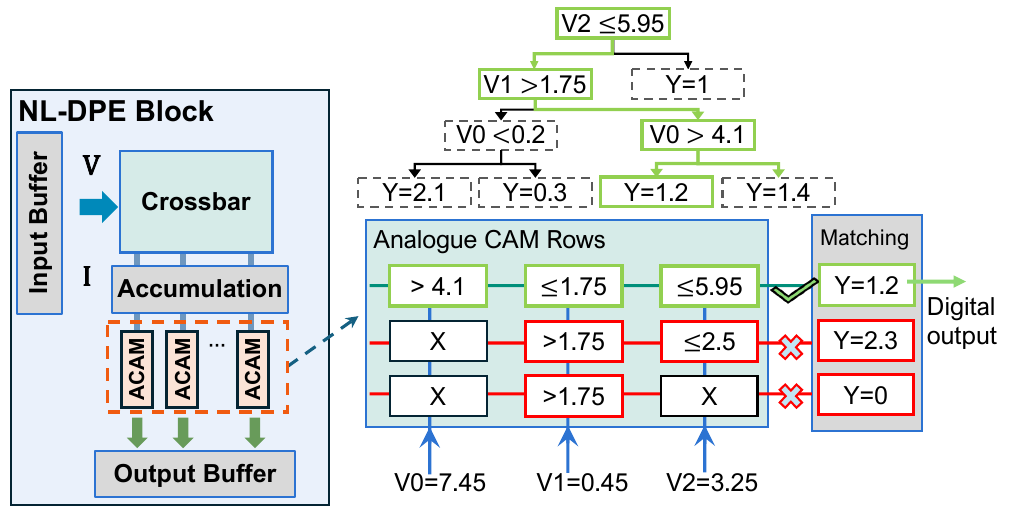}
\caption{\small Left: Block-level architecture of \textit{NL-DPE} showing ReRAM crossbar and ACAM units. Right: An example trained decision-tree mapped to the ACAM unit.}
\label{fig:nl_dpe_block}
\vspace{-0.3cm}
\end{figure}

Fig.~\ref{fig:nl_dpe_block} shows the block-level architecture of the \textit{NL-DPE} block. 
Each block contains a ReRAM crossbar of size $R{\times}C$, input/output buffers, and an array of $C$ ACAM units that replace the conventional ADC peripheral. Unlike the \textit{Azure-Lily} IMC block, which uses a single ReRAM cell per weight, each weight in the NL-DPE crossbar is encoded with four ReRAM cells to directly support signed MAC operations. This, however, does not increase the area of the overall crossbar significantly as the ReRAM cells contribute a small fraction of overall crossbar area.
For weight-persistent VMM operations, the crossbar computes VMM in the same manner as \textit{Azure-Lily}, but accumulated in the analog domain through ratioed capacitors before entering the ACAM. The ACAM maps the analog inputs to digital values while simultaneously applying a nonlinear activation function (ReLU, tanh, etc.).

The ACAM is itself a small ReRAM array whose cells are programmed with trained thresholds encoding a piecewise decision tree as shown in fig.~\ref{fig:nl_dpe_block}. Each ACAM unit is connected to one crossbar column and accepts analog input and performs a nearest-neighbor search over its programmed thresholds. In the figure, the ACAM weights on each row represent the threshold values stored in the decision tree. All the ACAM units together form a grid-style content-address memory, which produces the digital output value based on the input. By training different sets of ACAM thresholds, the same hardware can be reconfigured to achieve different nonlinear functions, which is analogous to how different crossbar weights implement different linear layers. It is also possible to program the ACAM units such that they only map analog inputs to digital values and not perform a nonlinear function. This programmability is the key enabler of performing in-block nonlinear functions. Both ACAM and crossbar weights are static during the application runtime and are only configured offline along with the bitstream generation.

While we focus on DL workloads in this paper, the NL-DPE can be used for other non-DL workloads as well. E.g., the crossbar's dot-product capability can be used for digital signal and image processing (FIR, FFT, Convolution). The pattern-matching capability of ACAM can be used for routing-table lookup and packet classification.

\subsection{Noise \& Precision Modeling}


Like all analog in-memory computing, both the ReRAM crossbar and the ACAM are subject to device noise, primarily from imprecise conductance programming and read-time conductance drifts. To compensate, we adopt the noise-aware finetuning (NAF) methodology of~\cite{nldpe,zhao_noise_finetune}, which jointly optimizes the crossbar weights and ACAM thresholds against a hardware-calibrated noise model. NAF is performed entirely in software prior to bitstream generation and requires no per-device calibration after deployment. At INT8 precision, it restores accuracy to within a few percent of the FP32 baseline across both CNN and Transformer workloads, with essentially no loss on BERT-Tiny~\cite{nldpe}.

\begin{table}[t]
\caption{IMC Block Interface }
\label{tab:imc_interface}
\centering
\scriptsize
\setlength{\tabcolsep}{3pt}
\renewcommand{\arraystretch}{0.95}
\begin{tabular}{@{}lcl@{\hskip 7pt}lcl@{}}
\toprule
\multicolumn{3}{c}{\textbf{Inputs}} & \multicolumn{3}{c}{\textbf{Outputs}} \\
\cmidrule(lr){1-3}\cmidrule(lr){4-6}
\textbf{Signal} & \textbf{Bits} & \textbf{Desc.} & \textbf{Signal} & \textbf{Bits} & \textbf{Desc.} \\
\midrule
clk              & 1  & Clock          & data\_out               & 40 & Out vector   \\
reset            & 1  & Reset          & dpe\_done               & 1  & Op.\ done    \\
data\_in         & 40 & In vector      & MSB\_SA\_ready          & 1  & MSB ready    \\
dpe\_ctrl        & 2  & DPE mode       & reg\_full               & 1  & Output full  \\
shift\_add\_ctrl & 1  & En.\ accum.    & shift\_add\_done        & 1  & Accum.\ done \\
shift\_add\_bypass & 1 & Byp.\ accum.  & shift\_add\_bypass\_ctrl & 1 & Byp.\ status \\
w\_buf\_en       & 1  & En.\ in.\ buf. &                         &    &              \\
load\_input\_reg & 1  & Load in.\ reg  &                         &    &              \\
load\_output\_reg & 1 & Latch output   &                         &    &              \\
\bottomrule
\end{tabular}
\end{table}

\subsection{IMC block interface and configuration}

The IMC block exposes a set of control, data, and status signals for coordinating data movement and execution. Table \ref{tab:imc_interface} summarizes the interface with brief functional descriptions. 
The 40-bit data interface on the IMC block is specifically chosen to match the widest BRAM configuration in our FPGA architecture (similar to Intel FPGA architectures). Unlike \textit{Azure-Lily}, which uses a 16-bit data interface, this enables us to exploit the maximum possible memory bandwidth for data transfer between BRAM and IMC block. 
The 40-bit data interface does not necessarily mean a compute precision of 40 bits. For example, in our evaluations using INT8 precision, we pack 5 elements in a single input vector thereby reducing the data transfer time by 5x.

In the proposed architecture, weights and decision thresholds are persistently stored within the ReRAM crossbar and ACAM units, respectively. Programming these non-volatile elements is analogous to initializing BRAM contents in a  baseline FPGA at boot time.
Outside the IMC block hierarchy, the standard FPGA configuration mechanism remains unchanged. 
Instead, the IMC blocks embed dedicated write logic and programming circuitry that interface directly with the configuration chain. This localized configuration logic uses the standard bitstream to program the ReRAM cells, ensuring that the weights and ACAM thresholds are loaded seamlessly as part of the overall chip configuration process. Consequently, this approach requires no invasive changes to the global FPGA configuration circuit or the conventional bitstream generation toolchain. Programming ReRAM cells also requires 
voltages and precise compliance currents that exceed the nominal FPGA core logic voltage. Consequently, integrating these macros requires dedicated internal power rails to support the programming phase. A detailed physical design and evaluation of this power delivery network, however, remain beyond the scope of this paper.



\subsection{FPGA-aware Design Space Exploration}

\begin{figure}[t]
\centering
\includegraphics[width=\columnwidth]{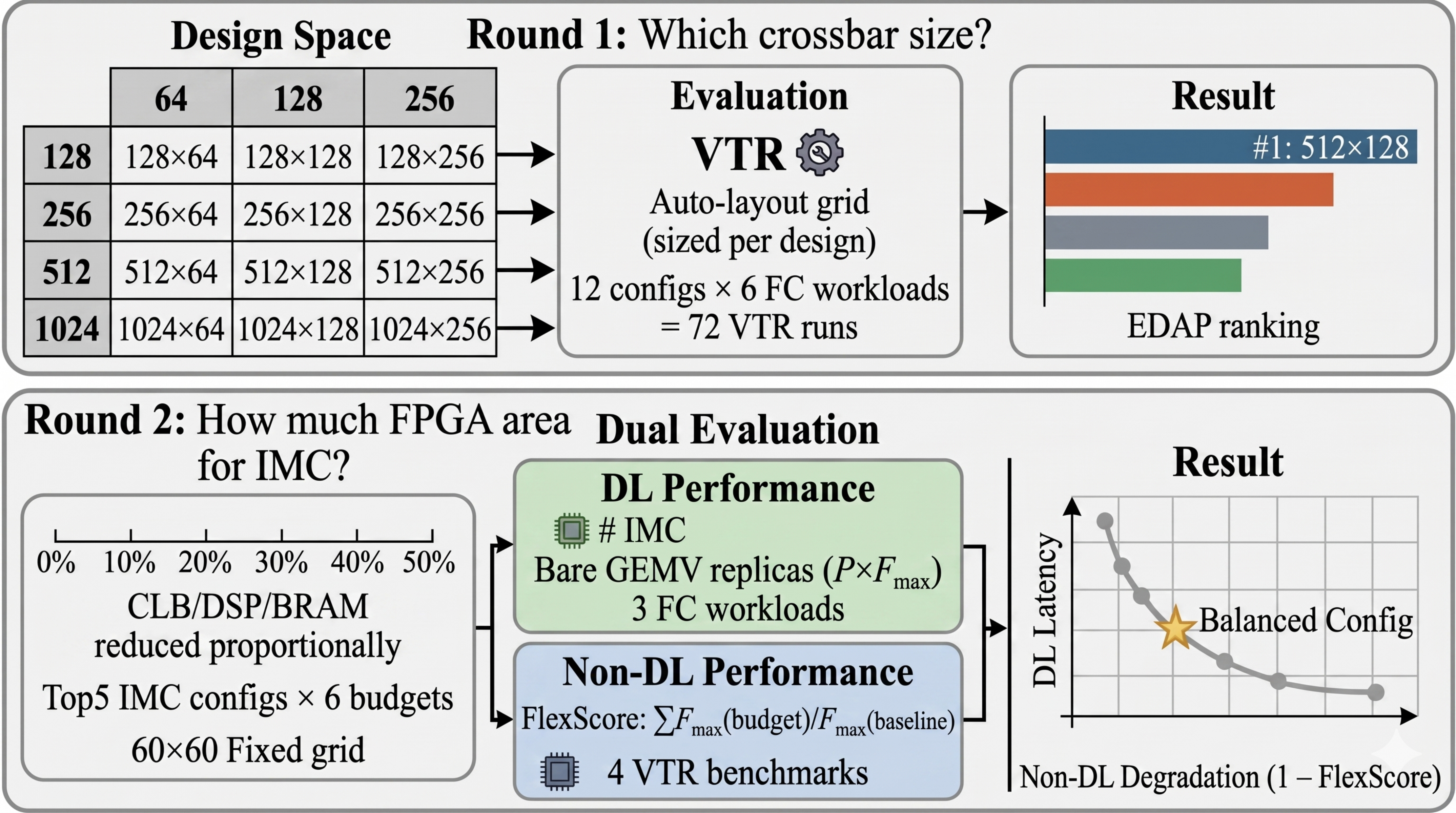}
\caption{\small Overview of the two-round FPGA-aware DSE. Round~1 selects the best crossbar sizes based on \textit{EDAP}. Round~2 sweeps IMC area budget on a fixed FPGA grid, balancing the DL throughput and flexibility.}
\label{fig:dse_overview}
\vspace{-0.3cm}
\end{figure}

We perform an FPGA-aware DSE to find the IMC configuration that best balances DL performance and FPGA flexibility. The DSE varies two inputs, the IMC crossbar size ($R{\times}C$) and the fraction of FPGA area allocated to IMC blocks, and evaluates efficiency, throughput, and architectural flexibility. We organize the DSE into two sequential rounds as shown in Fig.~\ref{fig:dse_overview}, where Round~1 selects the crossbar size and Round~2 sweeps the area budget.

\textbf{Round~1} evaluates the area, energy and delay tradeoff of different IMC crossbar configs. 
A crossbar with $R$ rows and $C$ columns holds a weight matrix of $R\times C$ and each column has an ACAM attached. 
The row count $R$ governs in-block activation eligibility: a layer of input dimension $M \times K$ maps to a single block when $K \leq R$, allowing ACAM to operate in activation mode and eliminating CLB activation overhead. When $K > R$, multiple IMC blocks are needed and the partial sums are reduced using CLBs. In this case, ACAM falls back to ADC mode, and activation must be handled in CLBs. At the same time, the column count $C$ governs horizontal tiling and per-block area, as the ACAM units count scales linearly with $C$ and each is significantly larger than crossbars. These effects create a non-trivial tradeoff: larger $R$ extends in-block activation to more layers but increases block area; larger $C$ reduces horizontal tiling but incurs significant area cost. 
Furthermore, a large crossbar will be underutilized for small VMM workloads, whereas small crossbars may require more soft logic to reduce results from multiple crossbars, resulting in another area-performance tradeoff.
Round~1 ranks candidate sizes on representative fully-connected layer (FC) workloads by \textit{Energy-Delay-Area Product} (\textit{EDAP}).

\textbf{Round~2} determines the IMC area budget, i.e. the percentage of area of the FPGA spent on IMC blocks. On a fixed FPGA grid, IMC columns are progressively substituted for CLB, BRAM, and DSP columns. Each budget point is evaluated along two axes simultaneously: (1)  Throughput on DL benchmarks, which increases with increasing IMC area, and (2) \textit{FlexScore} (based on \cite{flexscore}), the normalized frequency degradation on non-DL benchmarks caused by the removal of general FPGA resources. The resulting throughput--flexibility Pareto front identifies balanced area budget for each IMC configuration.

\subsection{Attention-Head Mapping}
\label{sec:mapping}
\begin{figure}[t]
\centering
\includegraphics[width=\columnwidth]{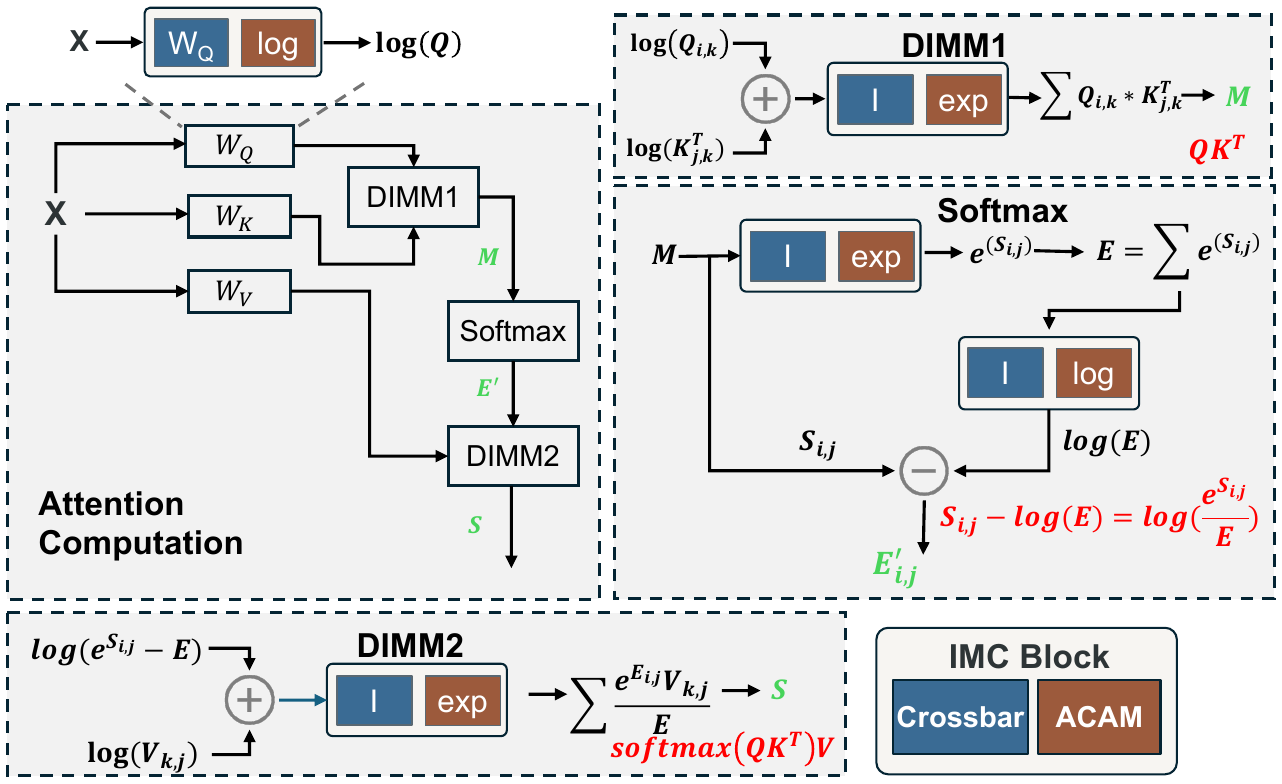}
\caption{\small Mapping strategy for Attention computation using the proposed IMC-based FPGA.}
\label{fig:k_identity_mapping}
\vspace{-0.3cm}
\end{figure}

Weight-persistence is a common approach used for FPGA-based DL inference \cite{brainwave}.
In this method, the pre-trained (aka static) weights of layers such as convolutional layers and fully-connected layers are stored on-chip to avoid external DRAM transfers.
These layers translate to VMM operations and can be accelerated by the IMC core by storing the weights in the crossbar during configuration time. 
However, 
Transformer-based workloads pose a new challenge for IMC-enhanced FPGAs.
The Attention mechanism in Transformers requires DIMM, $QK^T$ and score${\times}V$, whose operands are not static. Furthermore, \textit{softmax} and \textit{layernorm} introduce nonlinear operations between every Attention stage. In prior work, these DIMM and nonlinear stages rely entirely on soft logic, limiting the benefit of embedding more IMC blocks. In NIFA, however, we leverage the ACAM, to convert expensive MAC operations in DIMM to cheaper additions in the log domain, further extending the throughput and energy gains of IMC integration to the full Attention pipeline.

Fig.~\ref{fig:k_identity_mapping} illustrates the Attention head mapping onto the proposed FPGA. Linear Q/K/V projections use IMC crossbars with ACAM configured in log mode, producing log-domain outputs. DIMM stages ($QK^T$, score${\times}V$) operate entirely in the log domain. 
In the log domain, multiplications are replaced by additions (mapped to CLBs). 
In the IMC blocks used for DIMM stages, the crossbar is configured as an identity matrix that buffers the input and only performs the nonlinear functions through ACAM. 
Hence, the IMC block's outputs are in the linear domain and are then reduced to the final results using CLBs. 
Softmax is computed using multiple IMC blocks with ACAMs configured for either exp or log operations, as well as CLB based operations (addition and division converted to subtraction) as shown in the figure.

Using this mapping, the IMC blocks are reused at every stage of the Attention pipeline rather than falling back to DSPs and CLBs as in prior work, yielding significant performance gains. However, this mapping also raises a numerical-accuracy concern. As modeled in~\cite{nldpe}, a single transform is essentially exact at INT8, with a per-transform mean-squared error (MSE) on the order of $10^{-8}$, so individual transforms are not the concern. Error accumulates only when transforms are chained. A full log-domain multiply reaches an MSE of $10^{-5}$, and naively cascading stages would place the exponentiation and logarithm back-to-back that further amplifies the error. We therefore fuse these inverse stages so that they cancel rather than compound. At INT8, the log-domain DIMM and Softmax closely track their full-precision counterparts, with no measurable accuracy loss on BERT-Tiny\cite{nldpe}.

\vspace{-2mm} 
\section{Methodology}

\subsection{Tools Used}

\begin{figure}[t]
\centering
\includegraphics[width=\columnwidth]{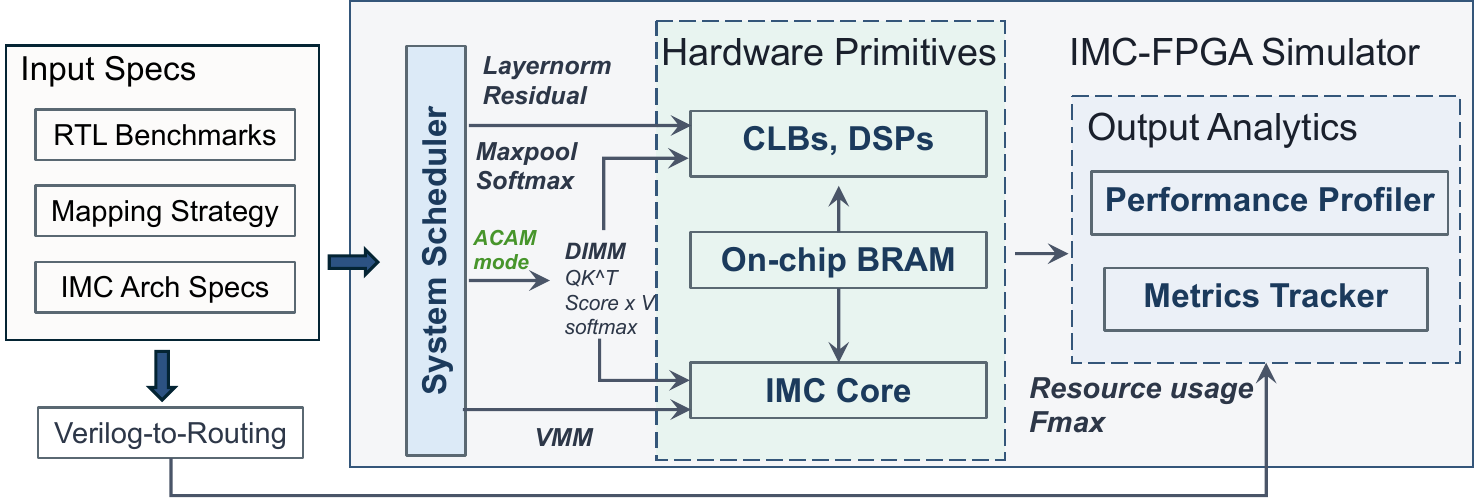}
\caption{\small Overview of our analytical simulator. VTR-reported Fmax and resource counts are combined with the energy model to produce per-layer latency and energy estimates.}
\label{fig:simulator_overview}
\end{figure}

In this work, we use VTR \cite{vtr9} for FPGA frequency and area evaluation. We also build a simulator for latency and energy profiling. A block diagram of the simulator is shown in Fig.~\ref{fig:simulator_overview}.
The mapping strategy for each end-to-end benchmark, informed by its RTL implementation, is input to the simulator, along with the specifications of the IMC block (such as rows, columns, area, energy).
A scheduling block follows this mapping strategy and partitions the layers across three compute paths: weight-persistent GEMMs on IMC blocks, 
DIMM operations ($QK^T$, softmax, score$\times V$) on both the IMC block and CLBs,  
and other operations such as \textit{LayerNorm}, \textit{Residual}, \textit{Maxpooling} on the CLBs. 
A performance profiler tracks inter- and intra-layer overlap, and a metrics tracker produces the total energy and end-to-end latency by aggregating per-layer metrics. 
The simulator models both ACAM and ADC based IMC blocks.
Each benchmark is implemented in Verilog and synthesized through VTR to obtain Fmax and resource counts, which are fed into the simulator as well. 
The energy calculation in our simulator includes three components: IMC energy, FPGA soft logic energy, and routing energy. 
Our simulator models energy consumption of all the components in the IMC block including input buffer, crossbar, ACAM, and output buffer. For the FPGA soft logic energy and routing energy, we use the analytical model from \cite{comefa_trets}, which estimates energy from logic resource usage and total routed wirelength.

\subsection{FPGA Architecture}

We use the 22\,nm Agilex-like FPGA 
architecture from \cite{koios}, the same architecture used by \textit{Azure-Lily}~\cite{azurelily}, as a baseline and augment it with our proposed IMC blocks.
To model the IMC area in VTR, we estimate the ReRAM crossbar area and ACAM area from a 32 nm NL-DPE design evaluated in \cite{nldpe}, and then scale the results to 22 nm using technology-scaling coefficients reported in \cite{stillmaker}.
Because the IMC block requires a relatively small number of input and output ports, its footprint is logic-bound rather than I/O-bound. Consequently, the area overhead of the local routing crossbar within the IMC tile is negligible compared to the core logic area.
To integrate these large hard blocks into the VTR grid, we model the IMC block to span multiple standard FPGA tiles in both width and height.
To further reduce the impact of integrating large hard blocks on the FPGA routing fabric, we preserve complete switch boxes at all the intersections where routing channels cross through the tile.
All inputs and outputs of the IMC block are registered and modeled with a 15\% input connection flexibility (FC\_in) and a 10\% output connection flexibility (FC\_out).

\subsection{Benchmarks}

\subsubsection{DL workloads for DSE} For DL throughput evaluation, six GEMV-style FC layers are used, as summarized in Table~\ref{tab:fc_workloads}. In addition to the core GEMV computation, each workload includes an activation function. The varied FC sizes capture the area-efficiency tradeoff across crossbar configurations: larger crossbars consume more area but can perform activation in-block, while smaller crossbars must pay the additional cost of CLB-based activation. The FC dimensions are drawn from representative CNN and Transformer layers, providing a realistic measure of crossbar utilization across workload types.

\begin{table}[t]
\centering
\caption{\small FC workloads and their representative DNN origins.}
\label{tab:fc_workloads}
\renewcommand{\arraystretch}{1.1}
\begin{tabular}{l|c|c|l}
\toprule
\textbf{Workload} & $K$ & $N$ & \textbf{Represents} \\
\midrule
fc\_64$\times$64     & 64   & 64  & Early CNN layers, tiny FC \\
fc\_128$\times$128   & 128  & 128 & Attention projection (Q/K/V) \\
fc\_512$\times$128   & 512  & 128 & ResNet mid-depth conv range \\
fc\_2048$\times$256  & 2048 & 256 & Deep CNN layers (ResNet, VGG) \\
fc\_256$\times$512   & 256  & 512 & Transformer FFN, VGG conv5+ \\
fc\_512$\times$512   & 512  & 512 & Large FC, Transformer projection \\
\bottomrule
\end{tabular}
\vspace{-0.2cm}
\end{table}
  
\subsubsection{Non-DL workloads for DSE}
Four non-DL designs are selected from the VTR benchmark suite: \textit{bgm}, \textit{LU8PEEng}, \textit{stereovision1}, and \textit{arm\_core}.  \textit{bgm} is CLB-intensive stressing routing and logic density; \textit{LU8PEEng} is BRAM-heavy representing memory-bound linear algebra; \textit{stereovision1} is DSP-dominant exercising compute-intensive pipelines; and \textit{arm\_core} combines CLBs with BRAMs as a general-purpose processor core. This selection ensures the evaluation covers all major FPGA resource types, preventing the \textit{FlexScore} from being biased toward any single resource profile.  As the IMCs are not utilized in these benchmarks, adding more IMCs will reduce other FPGA resources leading to higher routing congestion, the achievable Fmax therefore reflects the flexibility cost due to IMC integration.

\subsubsection{End-to-end benchmarks}For CNN evaluation, we use ResNet-9 and VGG-11, the same benchmarks used by \textit{Azure-Lily}, enabling direct comparison. For Transformer evaluation, we use BERT-Tiny (2 layers, 2 Attention heads, 128 hidden dimension, 512 FFN intermediate). All benchmarks follow the weight persistent methodology which is a common method for DNN deployment on FPGAs \cite{brainwave}.

\subsection{Metrics}
 \subsubsection{DSE}In Round~1, we rank crossbar sizes by \textit{EDAP}, aggregated via normalized geometric mean across workloads in Table \ref{tab:fc_workloads} (per-workload best $= 1.0$). In Round~2, we plot a Pareto-front to evaluate the tradeoff between DL workload throughput and architectural flexibility at each IMC area budget. The architectural flexibility is measured by \textit{FlexScore} \cite{flexscore}. The \textit{FlexScore} measures how much IMC hard blocks degrade non-DL workloads performance where IMC blocks are not used. 
 Each non-DL benchmark is synthesized through VTR and its Fmax is recorded. Each benchmark's Fmax is normalized to its own zero-IMC baseline — for example, \textit{bgm} achieves 90 MHz vs its baseline 100 MHz, giving a ratio of 0.9. This ratio is the \textit{Flexscore}. The overall \textit{FlexScore} of each architecture at each area budget is the geometric mean across the benchmarks. A geomean \textit{flexscore} of 0.96  means the non-DL workloads retain 96\% of their baseline Fmax under this given area budget and crossbar configuration.

\subsubsection{Non-DSE} For non-DSE experiments, we implement complete CNN models in RTL, along with BERT-Tiny models across all evaluated sequence lengths. All reported metrics are derived from VTR and our simulator. Energy is computed analytically via the simulator in \textit{pJ}. Area is reported by VTR in Minimum Width Area Transistors (MWTA) and is converted to $mm^2$. System latency is modeled by the simulator and reported in \textit{ns}. We further derive three composite efficiency metrics using these base metrics for system-level comparison including: inferences per second (throughput), throughput per mm$^2$, and inferences per joule.

\begin{figure}[t]
    \centering
    \includegraphics[width=0.9\columnwidth]{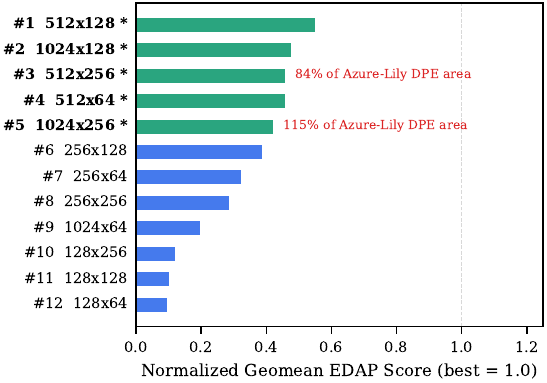}
    \caption{\small Round~1 DSE: Crossbar sizes ranked by \textit{EDAP}.}
    \label{fig:round1_ranking}
\end{figure}

\begin{table}[t]
\centering
\caption{\small IMC hard block comparison across three evaluation architectures at 22nm tech node}
\label{tab:dpe_block_comparison}
\renewcommand{\arraystretch}{1.15}
\footnotesize
\begin{tabular}{l c c c}
\toprule
\textbf{Property} & \textit{Proposed-1} & \textit{Proposed-2} & \textit{Azure-Lily} \\
\midrule
Crossbar ($R \times C$)       & 1024$\times$128  & 1024$\times$256  & 512$\times$128  \\
ReRAM Cells per weight                  & 4                & 4                & 1               \\
ACAM Size / ADC Count              &130$\times$128 &  130$\times$256 & 8 ADCs  \\
I/O data-width & 40-bit & 40-bit & 16-bit \\
Area (mm$^2$)           & 0.047            & 0.091            & 0.079           \\
FPGA grid size (rows$\times$cols)           & 3$\times$7            & 5$\times$8            & 6$\times$5           \\
Power (mW)              & 27.4             & 51.8             & 20.0            \\
TOPS(int8)              &16.4              & 32.8             & 0.91 \\
Frequency(GHz) &      0.9      &   0.9    &   0.92\\
\bottomrule
\end{tabular}
\vspace{-0.1cm}
\end{table}

\begin{figure}[]
    \centering
    \includegraphics[width=\columnwidth]{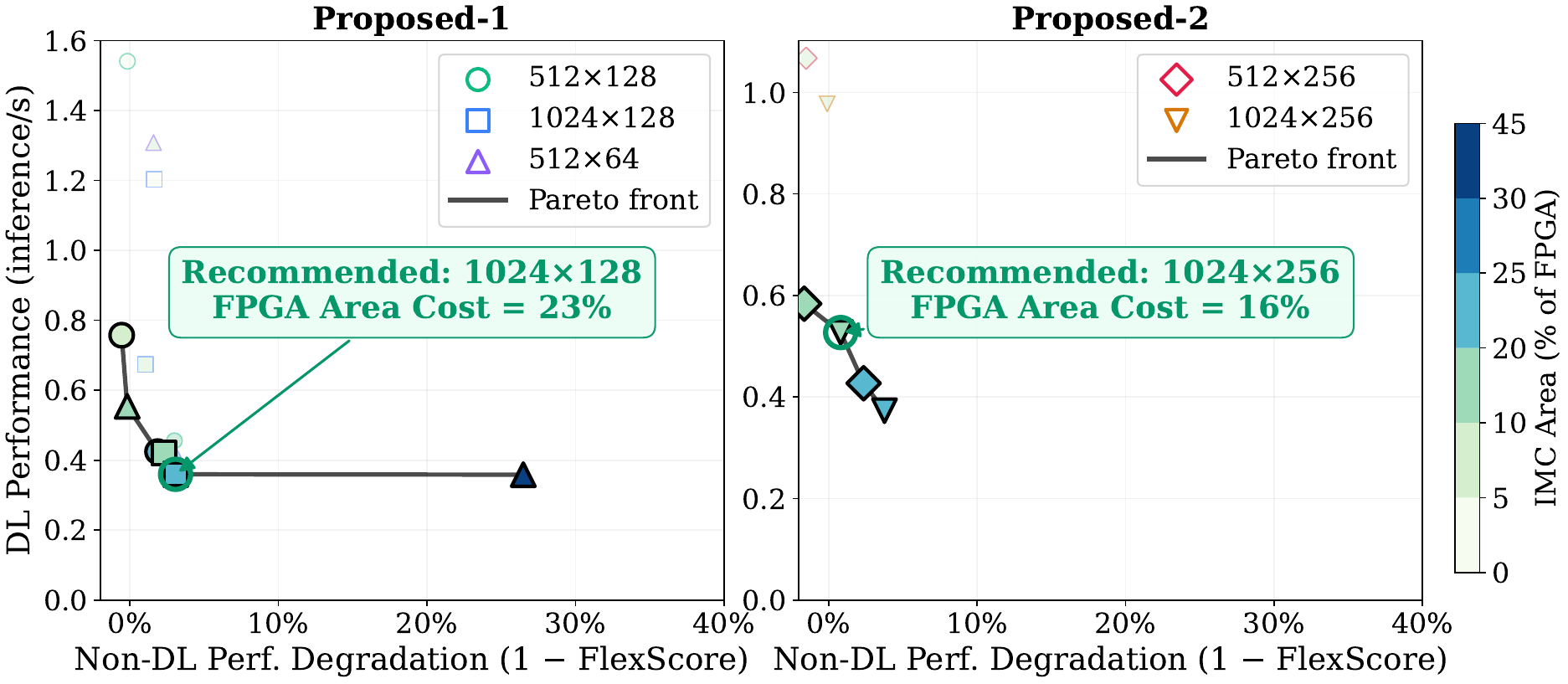}
    \caption{\small Round~2 DSE: Pareto-front evaluation of recommended crossbar sizes for \textit{Proposed-1} and \textit{Proposed-2} across DL performance and FPGA flexibility.}
    \label{fig:round2_pareto}
    \vspace{-0.2cm}
\end{figure}

\subsection{DSE Protocol}

\subsubsection{Round~1: Block Sizing.} We sweep 12 crossbar configurations ($R \in \{128, 256, 512, 1024\}$, $C \in \{64, 128, 256\}$) across all 6~FC workloads using VTR auto-layout. 
Each configuration is ranked by \textit{EDAP} geomean across benchmarks. From the results, we select the top-5 candidates for Round~2.

\subsubsection{Round~2: FPGA Integration.} We fix the FPGA size and sweep the IMC area budget by progressively replacing CLB, DSP, and BRAM tiles proportionally by IMC tiles — all three resource types lose the same fraction of capacity. 
Both DL throughput and \textit{FlexScore} are measured, producing a Pareto-front that identifies an appropriate balance of the FPGA area consumed by IMC resources vs. other FPGA resources.

\vspace{-2mm} 
\section{Results}



\subsection{Recommended Crossbar Sizes from DSE}

Fig.~\ref{fig:round1_ranking} ranks 12~crossbar configurations by \textit{EDAP} across the 6~FC workloads. A clear tier emerges: the top-5 including 512$\times$128, 1024$\times$128, 512$\times$256, 512$\times$64, and 1024$\times$256 all have $R \geq 512$, enabling in-block activation on the majority of workloads and eliminating CLB activation overhead. 
The top-5 advance to Round~2, split into two groups evaluated with identical workloads: \textit{Group~1} contains configs \#1, \#2, and \#4, and \textit{Group~2} contains the two configs with area similar to \textit{Azure-Lily} enabling a controlled block-level comparison.

Fig.~\ref{fig:round2_pareto} shows the Round~2 throughput--flexibility Pareto fronts. 
From these, we select two operating points for end-to-end evaluation: \textit{Proposed-1} (1024$\times$128, 23\% FPGA area) which achieves the best DL throughput at under 5\% flexibility degradation and \textit{Proposed-2} (1024$\times$256, 21\% FPGA area), a configuration whose IMC block area is comparable to \textit{Azure-Lily}'s IMC block, enabling a controlled comparison that shows our advantage at similar silicon cost. 
Other points yield negligible throughput gain while flexibility degrades sharply, confirming the importance of the DSE.


\subsection{Block-Level Evaluation}

\begin{figure}[t]
\centering
\includegraphics[width=\columnwidth]{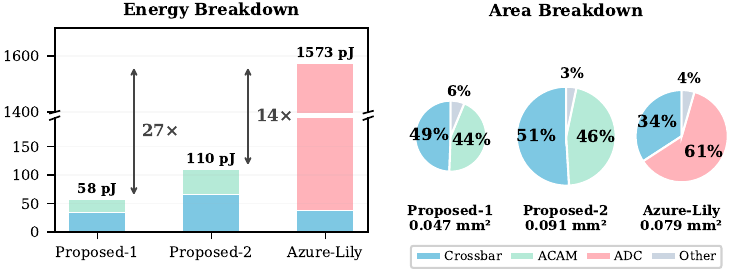}
\caption{\small Block-level energy and area comparison.}
\label{fig:block_comparison}
\end{figure}

Table~\ref{tab:dpe_block_comparison} summarizes the three IMC block configurations. 
Both proposed configs use $R{=}1024$ with a 40-bit data interface and 4 ReRAM cells per weight, enabling signed ACAM operations. \textit{Azure-Lily} uses $R{=}512$ with a 16-bit data interface and 1 cell per weight. 
Fig.~\ref{fig:block_comparison} compares the energy per VMM operation and the area breakdown of each IMC block. Notably, ADC energy dominates in \textit{Azure-Lily}, making \textit{Proposed-1} and \textit{Proposed-2} significantly more energy efficient while providing higher throughput due to their larger crossbar. In contrast, in the proposed designs, ACAM occupies an area comparable to the crossbar, whereas the ADC accounts for about 60\% of the block area in \textit{Azure-Lily}. This result indicates that ACAM provides both better area and energy efficiency than ADCs.


\begin{table}[t]
\centering
\caption{\small FPGA implementation results for CNN benchmarks.}
\label{tab:benchmark_resources_cnns}
\renewcommand{\arraystretch}{1.1}
\footnotesize
\setlength{\tabcolsep}{4pt}
\begin{tabular}{@{}llccc@{}}
\toprule
 & & \textit{Proposed-1} & \textit{Proposed-2} & \textit{Azure-Lily} \\
\midrule
\multirow{5}{*}{ResNet-9}
 & IMCs  & 19 (6\%)    & 12 (13\%)   & 35 (13\%) \\
 & DSPs  & 0           & 0           & 0         \\
 & CLBs  & 61 (0.4\%)  & 63 (0.4\%)  & 185 (2\%) \\
 & BRAMs & 28 (5\%)    & 28 (6\%)    & 16 (2\%)  \\
 & Fmax  & 168 MHz     & 158 MHz     & 215 MHz   \\
\midrule
\multirow{5}{*}{VGG-11}
 & IMCs  & 85 (29\%)   & 44 (49\%)   & 148 (57\%) \\
 & DSPs  & 0           & 0           & 0          \\
 & CLBs  & 99 (1\%)    & 101 (1\%)   & 212 (2\%)  \\
 & BRAMs & 30 (6\%)    & 30 (7\%)    & 12 (2\%)   \\
 & Fmax  & 133 MHz     & 138 MHz     & 154 MHz    \\
\bottomrule
\end{tabular}
\end{table}

\subsection{CNN Evaluation}

\begin{figure}[t]
\centering
\includegraphics[width=0.5\textwidth]{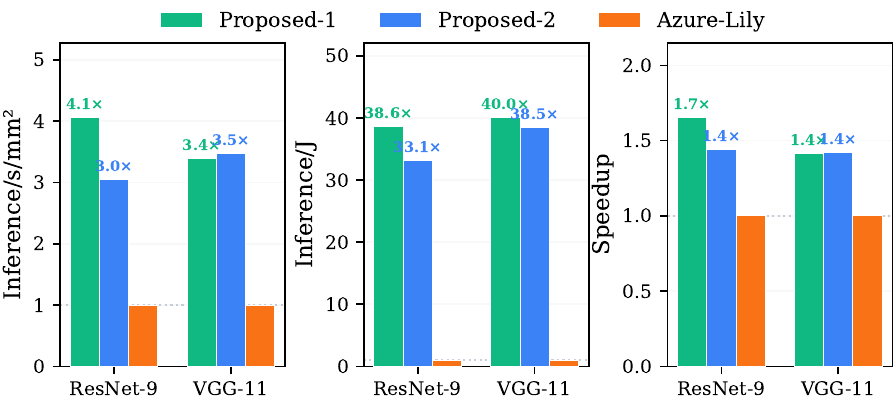}
\caption{\small CNN benchmark efficiency. Left: Area efficiency (Inference/s/mm$^2$). Middle: Energy efficiency (Inference/J). Right: Overall speedup}
\label{fig:cnn_efficiency}
\vspace{-0.2cm}
\end{figure}

Table~\ref{tab:benchmark_resources_cnns} reports the resource usage and Fmax obtained from VTR for each CNN benchmark. For these benchmarks, we use an FPGA grid size of 150x150 for all configurations.
Fig.~\ref{fig:cnn_efficiency} summarizes area, energy efficiency, and end-to-end speedup, all normalized to \textit{Azure-Lily}. Both proposed configurations achieve $3$--$4\times$ higher area efficiency on ResNet-9 and VGG-11. The improvement in energy efficiency is significantly larger: both \textit{Proposed-1} and \textit{Proposed-2} achieve over $30\times$ higher energy efficiency than \textit{Azure-Lily}.
This gain is partly due to ACAM eliminating the separate CLB-based activation passes required by \textit{Azure-Lily} after each IMC pass, leading to additional area and energy savings that may exceed the block-level energy differences shown in Fig.~\ref{fig:block_comparison}. Overall, the proposed configurations deliver a $1.4$--$1.7\times$ end-to-end speedup across ResNet-9 and VGG-11. Concretely, the CNN energy advantage stems from two block-level effects: replacing the ADC with the ACAM cuts the conversion energy by ${\sim}27\times$ (Fig.~\ref{fig:block_comparison}), and folding the activation into the ACAM removes the separate CLB activation pass that \textit{Azure-Lily} runs after every IMC pass. Together these account for the ${>}30\times$ end-to-end CNN energy efficiency over \textit{Azure-Lily}.

\subsection{BERT Evaluation}

\begin{table}[t]
\centering
\caption{\small FPGA implementation results for BERT-Tiny.}
\label{tab:benchmark_resources_bert}
\renewcommand{\arraystretch}{1.1}
\footnotesize
\setlength{\tabcolsep}{4pt}
\begin{tabular}{@{}llccc@{}}
\toprule
 & & \textit{Proposed-1} & \textit{Proposed-2} & \textit{Azure-Lily} \\
\midrule
\multirow{5}{*}{N=128}
 & IMCs  &  40  & 36   & 18  \\
 & DSPs  &  10  & 10   & 30  \\
 & BRAMs &  150 & 150  & 54  \\
 & CLBs  &  659 & 655  & 380 \\
 & Fmax  &  132 MHz   &  130MHz     &  131MHz    \\
\midrule
\multirow{5}{*}{N=2048}
 & IMCs  & 280 & 148  & 18  \\
 & DSPs  & 10  & 10  & 150        \\
 & BRAMs & 270 & 270  & 234   \\
 & CLBs  & 728 & 732  & 694 \\
 & Fmax  & 134 MHz     & 136 MHz     & 136 MHz    \\
\bottomrule
\end{tabular}
\end{table}

\begin{figure*}[!t]
\centering
\includegraphics[width=0.9\textwidth]{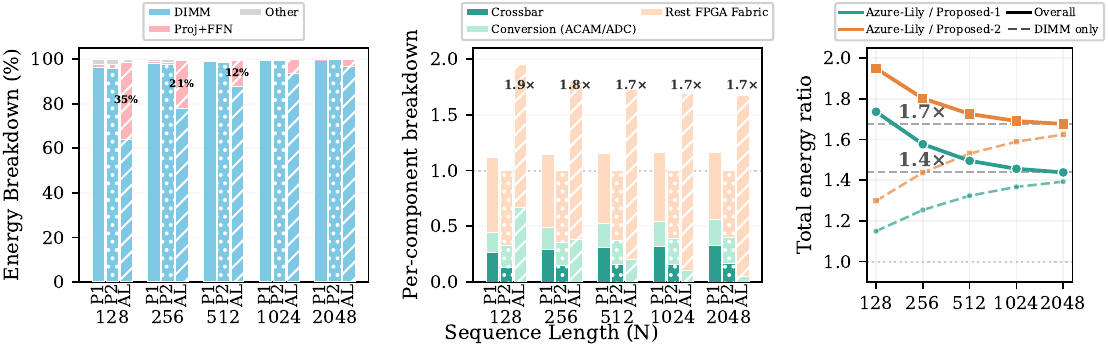}
\caption{\small BERT-Tiny energy analysis across sequence lengths. Left: Energy breakdown by operation. Middle: Energy breakdown by hardware component. Right: Total energy ratio of \textit{Azure-Lily} over \textit{Proposed-1} and \textit{Proposed-2}.}
\label{fig:bert_energy_3panel}
\end{figure*}

\begin{figure}[!tb]
\centering
\includegraphics[width=0.96\linewidth]{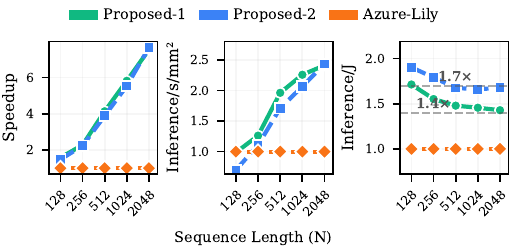}
\caption{\small BERT-Tiny speedup and efficiency (normalized to \textit{Azure-Lily}) across sequence lengths. Left: end-to-end speedup. Middle: Area efficiency. Right: Energy efficiency.}
\label{fig:bert_efficiency}
\vspace{-0.2cm}
\end{figure}

While the CNN results confirm the system-level benefits, the more distinctive contribution of NIFA lies in Transformer inference, where ACAM's nonlinear capabilities extend to activation, DIMM, and softmax stages. We conduct a detailed sensitivity study on BERT-Tiny across sequence lengths to evaluate how system-level benefits scale with the Attention head's $O(N^2)$ cost. 
Table~\ref{tab:benchmark_resources_bert} reports the resource usage and F\textsubscript{max} obtained from VTR for the largest and smallest sequence lengths evaluated for BERT-Tiny. All BERT-Tiny benchmarks are mapped onto an FPGA grid of 255$\times$255, the minimum grid size required to fit all designs. As described in Section~\ref{sec:mapping}, Attention head computation is accelerated by converting multiplications into additions in the log domain which results in notably higher CLB utilization for the proposed configurations compared to \textit{Azure-Lily}. Notably, all three architectures share the same LayerNorm implementation, which requires two multiplications per layer. Across the five LayerNorm layers in BERT-Tiny, this accounts for dedicated 10 DSPs.

We note that the proportion of DIMM operations within the total Attention FLOPs grows drastically with sequence length. From about 50\% at sequence length of 256 it approaches $100\%$ at sequence lengths $>=$ 4096. This trend makes it essential to efficiently accelerate the Attention mechanism as the model scales.


\subsubsection{Energy Breakdown}

Fig.~\ref{fig:bert_energy_3panel} reveals the source and scalability of the energy advantage across three panels. The left panel shows that DIMM operations dominate BERT-Tiny energy at every sequence length, with the non-DIMM share shrinking from 35\% at $N{=}128$ to 12\% at $N{=}512$, confirming DIMM as the system energy bottleneck. The middle panel decomposes per-element DIMM energy into crossbar, conversion peripheral, and fabric components. Two trends stand out: first, \textit{Azure-Lily}'s ADC-based conversion is far more expensive than ACAM; second, its fabric energy share is significantly larger because DIMM operations fall back to DSPs, whereas the proposed architectures accelerate DIMM in the log domain via ACAM and CLBs which consumes substantially less power than DSPs. Together, these factors produce an energy ratio that starts at $1.9\times$ and converges to a persistent ${\sim}1.7\times$ floor as $N$ grows, demonstrating that the advantage scales robustly with $O(N^2)$ Attention cost. 
The right panel confirms this: as DIMM dominates total energy, the system ratio converges to the DIMM ratio, with \textit{Azure-Lily} consuming ${\sim}1.4\times$ and ${\sim}1.7\times$ the energy of \textit{Proposed-1} and \textit{Proposed-2}.

\subsubsection{End-to-End Efficiency}
Fig.~\ref{fig:bert_efficiency} presents the speedup and efficiency normalized to \textit{Azure-Lily}. 
The left subfigure shows the end-to-end speedup against \textit{Azure-Lily}. As N grows, both proposed configurations deliver similar, consistently increasing speedup.  
Area efficiency (middle sub-figure) increases consistently with sequence length for both proposed configurations: \textit{Proposed-1} rises from $1.0\times$ at $N{=}128$ to ${\sim}2.4\times$ at $N{=}2048$, and \textit{Proposed-2} follows a similar trajectory. As $N$ grows, DIMM operations increasingly dominate total compute, and the ACAM-based IMC block's throughput advantage compounds accordingly. The monotonically increasing trend demonstrates strong scalability in both area utilization and sequence length.
Energy efficiency (right sub-figure) tells a consistent story: both configurations converge to stable floors: $1.4\times$ for \textit{Proposed-1} and $1.7\times$ for \textit{Proposed-2} as observed in the per-component energy analysis (Fig.~\ref{fig:bert_energy_3panel}). These results confirm that the proposed architecture's end-to-end efficiency advantage stands robustly as $O(N^2)$ Attention cost grows.

\vspace{-2mm} 
\section{Conclusion}

In this work, we present \textit{NIFA}, a heterogeneous FPGA architecture that integrates ADC-free, ACAM-based analog IMC blocks as first-class hard blocks, enabling native nonlinear computation inside the IMC core. A systematic two-round FPGA-aware DSE jointly optimizes the trade-off between FPGA area, DL throughput, and architectural flexibility, and an efficient mapping extends the IMC block's applicability to dynamic-input matrix multiplications in Transformer Attention operation. Across CNN and BERT-Tiny benchmarks, the proposed architecture demonstrates significant energy and performance efficiency over state-of-the-art analog IMC based FPGA architectures. 
Most importantly, this advantage holds as input sequence length grows at $O(N^2)$, confirming the architecture is efficient and scalable.


\vspace{-2mm} 
\section{Acknowledgements}
This work was supported in part by National Science Foundation (grant number 2417658). Any opinions, findings, conclusions, or recommendations are those of the authors and not of the funding institutions. 
The authors acknowledge the use of AI assistants to assist with manuscript drafting and to support code development and debugging. All core ideas, experimental designs, and scientific conclusions are entirely the original work of the authors.



%

\bibliographystyle{ieeetr}
\bibliography{bibfile}



\end{document}